\documentclass[twocolumn,showpacs,preprintnumbers,amsmath,amssymb]{revtex4}

\usepackage{CJK}
\usepackage{graphicx}
\usepackage{dcolumn}
\usepackage{bm}

\def\btbl{\begin{tabular}} \def\etbl{\end{tabular}}
\def\bcc{\begin{center}} \def\ecc{\end{center}}
\def\beq{\begin{equation}} \def\eeq{\end{equation}}
\def\btbl{\begin{tabular}} \def\etbl{\end{tabular}}

\def\E941{{\footnotesize E941}} \def\E864{{\footnotesize E864}}
\def\NA49{{\footnotesize NA49}} \def\NA35{{\footnotesize NA35}}

\begin{document}
\title{A systematic study of magnetic field in Relativistic Heavy-ion Collisions in the RHIC
and LHC energy regions }

\author{Yang Zhong$^{1,3,*}$}
\author{Chun-Bin Yang$^{1,2}$}
\author{Xu Cai$^{1,2}$}
\author{Sheng-Qin Feng$^{3,2}$}

\affiliation{$^1$ Institute of Particle Physics, Central China Normal University, Wuhan 430079, China}
\affiliation{$^2$ Key Laboratory of Quark and Lepton Physics (MOE), Central China Normal University, Wuhan 430079, China}
\affiliation{$^3$ Department of Physics, College of Science, China Three Gorges University, Yichang 443002, China}
\affiliation{$^*$Correspondence should be addressed to Yang Zhong; yzhong913@163.com }

\begin{abstract}
The features of magnetic field in relativistic heavy-ion collisions are systematically
studied  by using a  modified magnetic field model in this paper. The features of magnetic field distributions in the central point are studied
in the RHIC and LHC energy regions.  We also predict the
feature of  magnetic fields at LHC $\sqrt{s_{NN}}$= 900, 2760 and 7000 GeV based on
the detailed study at RHIC  $\sqrt{s_{NN}}$ = 62.4, 130 and 200 GeV. The dependencies of the features of magnetic fields on the
collision energies, centralities and collision time are systematically investigated, respectively. \\

\end{abstract}

 \maketitle

\section{Introduction}
\label{intro}
Collisions of two heavy nuclei at high energy serve as a
means for creating and exploring strongly interacting matter at
high possible energy densities where a new extreme state of matter,
the deconfined quark-gluon plasma (QGP) is expected to be formed ~\cite{lab1,lab2,lab3}.
Besides the formation of QGP, relativistic heavy-ion collisions create  also
extremely strong (electro)magnetic field due to the relativistic
motion of the colliding heavy ions carrying large positive electric
charge~\cite{lab3, lab4, lab5, lab6}.

Now we turn to a key question: can the Chiral Magnetic Effect(CME) occur in
heavy ion collisions? The answer to the  question seems to be positive. Two elements are
needed for the CME to occur: an external magnetic field and a locally nonzero
axial charge density. The relativistically moving heavy ions, typically with
large positive charges (e.g. +79e for Au), carry strong magnetic (and electric)
fields with them. In the short moments before/during/after the impact of
two ions in non-central collisions, there is a very strong magnetic field in the
reaction zone~\cite{lab5,lab6,lab7}. In fact, such a magnetic field is estimated to be of the
order of $m_{\pi}^{2}\approx 10^{18}$ Gauss ~\cite{lab4,lab8,lab9}, probably the strongest,
albeit transient, magnetic field in the present universe. The other required
element, a locally non-vanishing axial charge density, can also be created in
the reaction zone during the collision process through sphaleron transitions
(see e.g.~\cite{lab10} for discussions and references therein). As such, it appears at
least during the very early stage of a heavy ion collision, there can be both
strong magnetic field and nonzero axial charge density in the created hot
matter.

It is suggested by Ref.~\cite{lab5,lab6,lab11,lab12,lab13} that off-central heavy-ion collisions can create strong transient
magnetic fields due to the fast, oppositely directed motion of two colliding
ions.  Thus, heavy-ion collisions provide a unique terrestrial
environment to study QCD in strong magnetic fields~\cite{lab14,lab15,lab16,lab17}. It has
been shown that a strong magnetic field can convert topological
charge fluctuations in the QCD vacuum into global electric
charge separation with respect to the reaction plane. This so-called chiral magnetic effect may serve as a sign of the
local P and CP violation of QCD.

Experimentally, the STAR ~\cite{lab18,lab19,lab20,lab21}, PHENIX~\cite{lab22} and ALICE ~\cite{lab23} Collaborations have
presented the measurements of CME by the two-particle or three-particle correlations of
charged particles with respect to the reaction plane, which
are qualitatively consistent with the CME. A clear signal compatible with a charge dependent
separation relative to the reaction plane is observed, which shows little or no collision energy
dependence when compared to measurements at RHIC energies. This provides a new insight for understanding
the nature of the charge-dependent azimuthal correlations observed at RHIC and LHC energies.

Kharzeev, Mclerran and Warringa (KMW) ~\cite{lab5} suggested that in the vicinity
of the deconfinement phase transition, and under the influence of the strong magnetic field generated by the colliding
nuclei, the quark spin alignment along the direction of the angular momentum and the imbalance of the left- and right-handed quarks
generates an electromagnetic current. The experimental search of those effects has intensified recently, following
the realization that the consequent quark fragmentation
into charged hadrons results in a charge separation along
the direction of the magnetic field, and perpendicular to the
reaction plane.

The magnitude of the effect should either not change or should
decrease with increasing energy as long as a deconfined state of matter is formed in a heavy-ion
collision. In addition, in ~\cite{lab24,lab25} it
is also suggested that there should be no energy dependence
between the top RHIC and the LHC energies, based
on arguments related to the universality of the underlying
physical process, without however explicitly quantifying
what the contribution of the different values and time
evolution of the magnetic field for different energies will
be. On the other hand, it is argued~\cite{lab26} that the CME
should strongly decrease at higher energies, because the magnetic field decays more rapidly. Such a spread in the
theoretical expectations makes it important to measure the charge-dependent azimuthal correlations at the LHC,
where the collision energy is an order of magnitude higher compared to the RHIC.

Charge separation needs a symmetry axis along which the separation can take place. The
only symmetry axis in a heavy-ion collision is along the angular momentum which is
perpendicular to the reaction plane. In central collisions there is no symmetry axis, so in that case
charge separation should vanish. The strong magnetic field and the QCD vacuum can both completely be produced
in the non-central nucleus-nucleus collision.

In Ref.~\cite{lab27}, we used the Wood-Saxon nucleon distribution instead of
a uniform one to improve the magnetic field calculation of the magnetic field for non-central collision. It was argued that the magnitude of the magnetic field decreases at higher energies.
The detailed research of the magnetic field dependencies on collision energies, impact parameter and collision times is presented in this paper.

The paper is organized as follows. The improved calculation of  magnetic field and the comparison of our new results with
that given by KMW  are described in Sec. II, along with the predicted results of LHC energy region. The produced particle contribution
to the magnetic field is considered  in Sec. III.
A summary is given in Sec. IV.

\section{The Improved calculation of Magnetic field}
The situation with the experimental search for the local
strong parity violation drastically changed once it was
noticed ~\cite{lab28,lab29,lab30} that in noncentral nuclear collisions it would
lead to the asymmetry in the emission of positively
and negatively charged particle perpendicular to the reaction
plane. Such a charge separation is a consequence of
the difference in the number of quarks with positive and
negative helicities positioned in the strong magnetic field of a noncentral nuclear collision, the so-called
chiral magnetic effect.

We begin with a charged particle moving along the direction of the $z$ axis.
The magnetic field around it can be given by
\begin{equation}
\vec{B}=\frac{1}{c^2}\vec{v}\times\vec{E}
\label{eq:eq1} 
\end{equation}

If the movement is relativistic, at the time $t=0$, the charge is the origin of the coordinate. The magnitude of the magnetic field
$\vec{B}$ is given by
\begin{eqnarray}
B=\frac{1}{4\pi\varepsilon_0c^2}
\frac{qv(1-\beta^2)\sin\theta}{r^2(1-\beta^2\sin\theta)^{3/2}}.
\label{eq:eq2} 
\end{eqnarray}

Now we consider a particle with charge $Z$ and rapidity $Y$ traveling along the $z$ axis.
At $t=0$ the particle can be found at position $\vec{x}^{\prime}_{\bot}$; the magnetic field
at the position $\vec{x}=(\vec{x}_\bot,z)$ caused by the particle is given by

\begin{eqnarray}
\lefteqn{e\vec{B}(\vec{x})=Z\alpha_{EM}\sinh Y\times}\nonumber\\
&&\frac{(\vec{x}^{\prime}_{\bot}-\vec{x}_\bot)\times\vec{e}_z}
{[(\vec{x}^{\prime}_{\bot}-\vec{x}_\bot)^2+(t\sinh Y-z\cosh Y)^2]^{3/2}}
\label{eq:eq3} 
\end{eqnarray}

Now we suppose two similar nuclei with charge $Z$ and radius $R$ are traveling
in the positive and negative $z$ direction with rapidity $Y_0$.
At $t=0$ they have a noncentral collision with impact parameter $b$ at the origin point.
We take the center of the two nuclei at $x=\pm b/2$ at  time $t=0$
so that the direction of $b$ lies along the $x$ axis.

As the nuclei are nearly traveling with the speed of light in typical heavy-ion collision experiments,
the Lorentz contraction factor $\gamma$ is so large that we can consider the two included nuclei as pancake shaped.
As a result, the nucleon's number density of each nuclei at $\vec{x}^{\prime}=(\vec{x}^{\prime}_{\bot},z)$
can be given by

\begin{eqnarray}
\rho_{s\pm}(\vec{x}^{\prime}_{\bot})=\frac{2}{4/3\pi R^3}
\sqrt{R^2-(\vec{x}^{\prime}_{\bot}\pm\vec{b}/2)^2}
\label{eq:eq4} 
\end{eqnarray}

As a result, it seems that the nucleon distribution on average in a nucleus is an approximate result before considering
the Lorentz contraction. In Ref.[1], KMW model used the uniform nuclear distribution as the nuclear distribution. But for a real situation, the nucleon distribution is not strictly uniform.
It seems more reasonable to use the Wood-Saxon distribution in place of the uniform
distribution. We use the Wood-Saxon distribution in this paper,

\begin{eqnarray}
n_A(r)=\frac{n_0}{1+\exp{(\frac{r-R}{d})}},
\label{eq:eq5} 
\end{eqnarray}

\noindent here $n_0$=0.17fm$^{-3}$, $d$=0.54fm and the radius $R$=1.12A$^{1/3}$fm. Considering the Lorentz contraction,
the density in the two-dimensional plane can be given by:

\begin{eqnarray}
\rho_{\pm}(\vec{x}^\prime_\bot)=N\cdot\int_{-\rm R}^{\rm R}dz'\frac{n_0}{1+\exp(\frac{\sqrt{(x'\mp{b/2})^2+y'^{2}+z'^{2}}-{\rm R}}{d})},
\label{eq:eq6} 
\end{eqnarray}

\noindent where $N$ is the normalization constant. The number densities should be normalized as

\begin{eqnarray}
\int{d}\vec{x}^\prime_\bot\rho_{\pm}(\vec{x}^\prime_\bot)=1.
\label{eq:eq7} 
\end{eqnarray}

We now estimate the strength of the magnetic field at position $\vec{x}=(\vec{x}_\bot,z)$
caused by the two traveling nuclei. We are only interested in the time $t>0$, i.e. just after the collision.
Then we can split the contribution of particles to the magnetic field in the following way

\begin{eqnarray}
\vec{B}=\vec{B}^+_s+\vec{B}^-_s+\vec{B}^+_p+\vec{B}^-_p
\label{eq:eq8} 
\end{eqnarray}

\noindent where $\vec{B}^\pm_s$ and $\vec{B}^\pm_p$ are the the contributions of the spectators and the participants
moving in the positive or negative $z$ direction, respectively. For spectators, we assume that they do not scatter at all
and that they keep travelling with the beam rapidity $Y_0$. According to Eq.(3), we use the density above and find

\begin{eqnarray}
\lefteqn{e\vec{B}^\pm_s(\tau,\eta,\vec{x}_\bot)=\pm Z\alpha_{EM}\sinh(Y_0\mp\eta)
\int{d}^2\vec{x}^\prime_\bot\rho_{\pm}(\vec{x}^\prime_\bot)}\nonumber\\
&&\times[1-\theta_\mp(\vec{x}^\prime_\bot)]\frac{(\vec{x}^\prime_\bot-\vec{x}_\bot)\times\vec{e}_z}
{[(\vec{x}^\prime_\bot-\vec{x}_\bot)^2+\tau^2\sinh(Y_0\mp\eta)^2]^{3/2}},
\label{eq:eq9} 
\end{eqnarray}

\noindent where $\tau=(t^2-z^2)^{1/2}$ is the proper time, $\eta=\frac{1}{2}\ln[(t+z)/(t-z)]$ is the space-time rapidity, and

\begin{eqnarray}
\theta_\mp(\vec{x}^\prime_\bot)=\theta[R^2-(\vec{x}^\prime_\bot\pm\vec{b}/2)^2].
\label{eq:eq10} 
\end{eqnarray}

Here, we would like to  neglect the contribution from the particles created by the interactions
and so we just need to take into account the contribution of the participants that were originally there.
The distribution of participants that remain travelling along the beam axis is given by
\begin{eqnarray}
f(Y)=\frac{a}{2\sinh(aY_0)}{\rm e}^{aY},  \hskip1cm -Y_{0}\leq{Y}\leq{Y_{0}}
\label{eq:eq11} 
\end{eqnarray}

\noindent Experimental data shows that $a\approx1/2$, consistent with the baryon junction stopping mechanism. The contribution
of the participants to the magnetic field can be also given by

\begin{eqnarray}
e\vec{B}^\pm_p(\tau,\eta,\vec{x}_\bot)=\pm Z\alpha_{EM}\int{\rm d}^2\vec{x}^\prime_\bot
\int{\rm d}Y f(Y)\sinh(Y\mp\eta)\nonumber\\
\times\rho_{\pm}(\vec{x}^\prime_\bot)\theta_\mp(\vec{x}^\prime_\bot)
\frac{(\vec{x}^\prime_\bot-\vec{x}_\bot)\times\vec{e}_z}
{[(\vec{x}_\bot^\prime-\vec{x}_\bot)^2+\tau^2\sinh(Y\mp\eta)^2]^{\frac{3}{2}}}
\label{eq:eq12} 
\end{eqnarray}

We calculate the magnetite of the magnetic field at the origin ($\eta=0,\vec{x}_\bot=0$) in which case
it is pointing in the $y$ direction. We took a Au-Au collision with different beam rapidities
and different impact parameters.

\begin{figure}[h!]
\centering \resizebox{0.52\textwidth}{!}{
\includegraphics{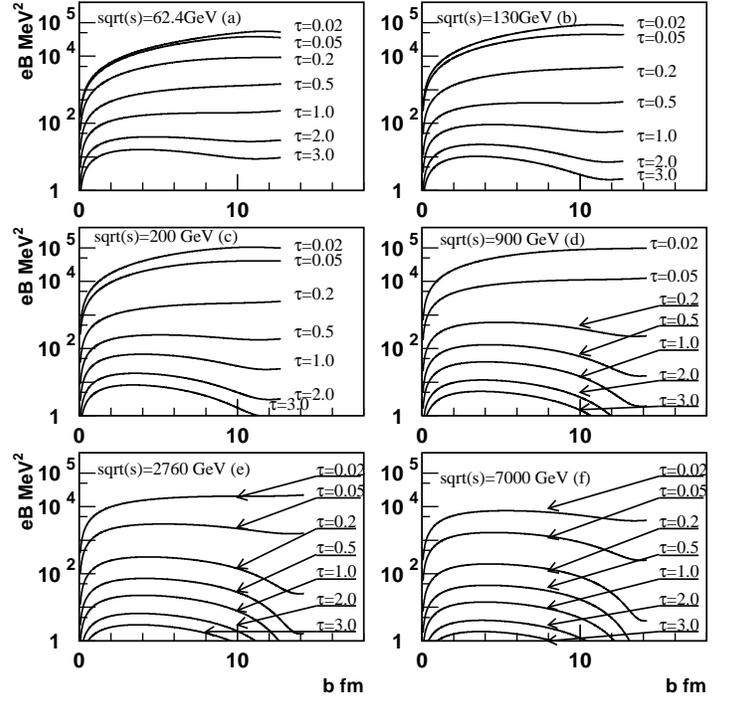}}
\caption{The dependencies of magnetic fields on impact parameters at different proper times $\tau$ at different energies of central mass system of
$\sqrt{s_{NN}}$=62.4 GeV(a), 130 GeV(b), 200 GeV(c), 900 GeV(d), 2760 GeV(e) and 7000 GeV(f), respectively.}
\label{fig1}
\end{figure}

Figure 1 shows that the dependencies of magnetic fields of the central point on impact parameters at different proper times $\tau$ at different center-of-mass energies for
RHIC and LHC, respectively. Fig.1(a, b and c) is for the results in RHIC energy region.  It is shown from Fig.1(a, b and c) that the magnitudes of the magnetic fields increase with the increasing impact parameter as proper time $\tau\leq1.0$ fm. Fig.1(d, e and f) is for the results in LHC energy region.  Fig.1(d and e) show that the magnitudes of the magnetic field increase with the increasing impact parameter only at proper time $\tau\leq0.05$ fm, but start to
decrease with increasing impact parameter when proper time $\tau>0.05$ fm. Fig. 1(f) shows that the magnitudes of the magnetic field decrease with the increasing impact parameter at proper time $\tau\geq0.02$ fm at $\sqrt{s_{NN}}$ = 7000 GeV.  Comparing with that of RHIC energy region, we find that the magnitudes of the magnetic fields of $\tau\geq2$ fm fall to zero more rapidly at LHC energy region. The variation characteristics of magnetic field with impact parameter at RHIC energy region are different from that of LHC energy region.

\begin{figure}[h!]
\centering \resizebox{0.5\textwidth}{!}{
\includegraphics{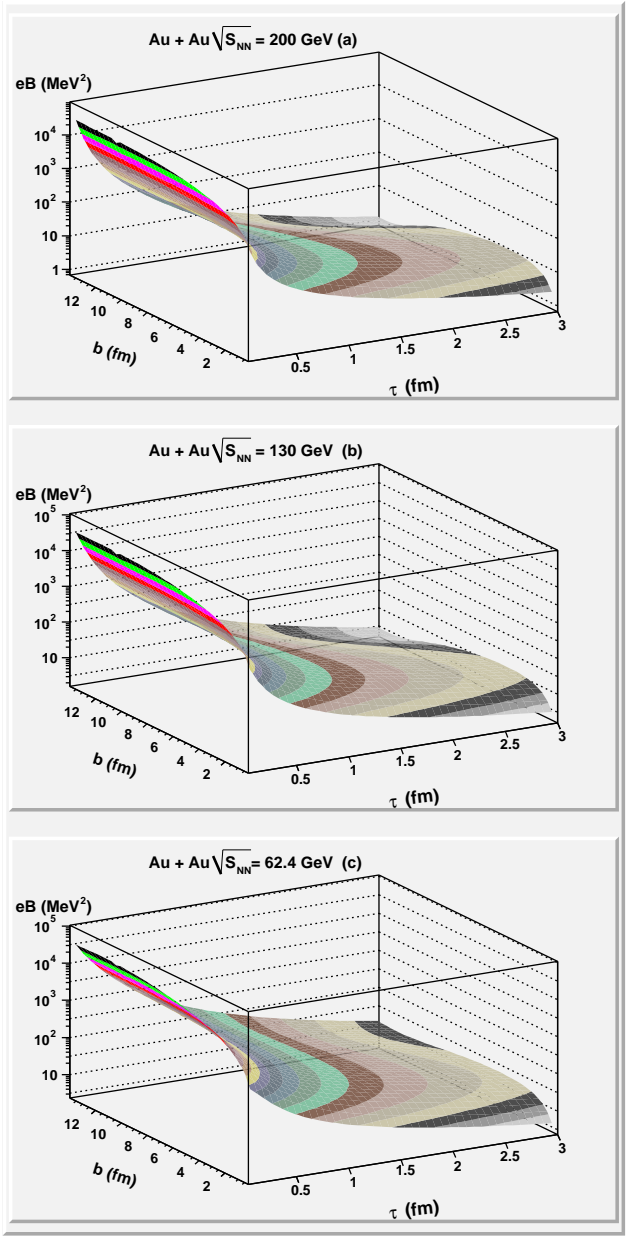}}
\caption{The dependencies of the magnetic field on proper time and impact parameters for
Au-Au collisions at RHIC collision energies with  $\sqrt{s}$ =  200 GeV(a), $\sqrt{s}$ =  130 GeV(b)
and $\sqrt{s}$ = 64 GeV(c), respectively.}
\label{fig2}
\end{figure}

The  dependencies of the magnetic field on proper time and impact parameters for
Au-Au collisions at RHIC energy region with  $\sqrt{s}$ =  200 GeV , $\sqrt{s}$ =  130 GeV
and $\sqrt{s}$ = 64 GeV are shown in Fig.2 (a, b, c), respectively.  The similar tendency of the dependencies of the magnetic field on proper time and impact parameters at RHIC
is observed from Fig.2. The maximum position of magnetic field is located at much small proper time $\tau\sim0.02 fm$ and much large impact parameter $b\simeq 12 fm$ shown in this Figure.
It is observed that the magnitude of magnetic field decreases sharply with increasing proper time $\tau$, and decreases with decreasing impact parameter $b$.

\begin{figure}[h!]
\centering \resizebox{0.5\textwidth}{!}{
\includegraphics{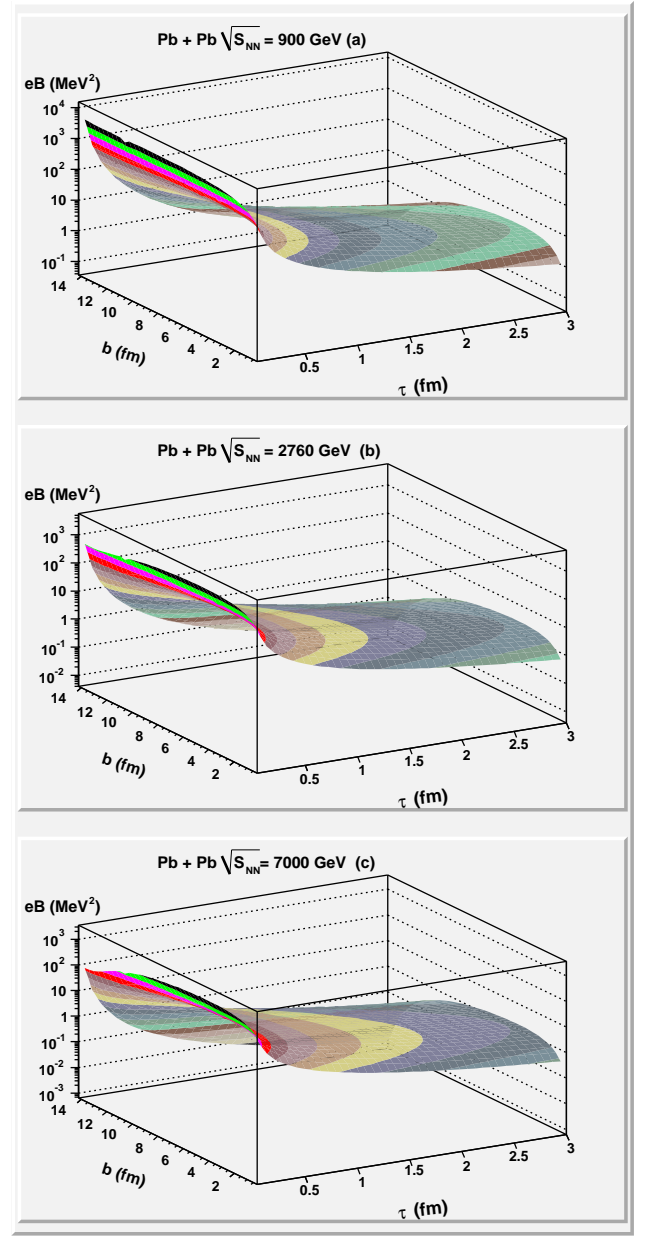}}
\caption{The dependencies of the magnetic field on proper time and impact parameters for
Pb-Pb collisions at LHC collision energies with $\sqrt{s}$ =  900 GeV(a), $\sqrt{s}$ =  2760 GeV(b)
and $\sqrt{s}$ = 7000 GeV(c), respectively.}
\label{fig3}
\end{figure}

Figure 3 shows the dependencies of the magnetic field on proper time and impact parameters for
Pb-Pb collisions at LHC collision energies with $\sqrt{s}$ =  900 GeV(a), $\sqrt{s}$ =  2760 GeV(b) and $\sqrt{s}$ = 7000 GeV(c), respectively. It is found that the maximum of magnetic field $eB\sim3.0\times10^{3}{\rm MeV}^{2}$ at  $\sqrt{s}$ =  900 GeV,
$eB\sim4.0\times10^{2}{\rm MeV}^{2}$ at  $\sqrt{s}$ =  2760 GeV, and $eB\sim80.0{\rm MeV}^{2}$ at  $\sqrt{s}$ =  7000 GeV. These are much smaller than that in
RHIC energy region. The similar variation characteristics of magnetic field with impact parameter and proper time are observed at LHC energy region and that
of RHIC.

\begin{figure}[h!]
\centering \resizebox{0.5\textwidth}{!}{
\includegraphics{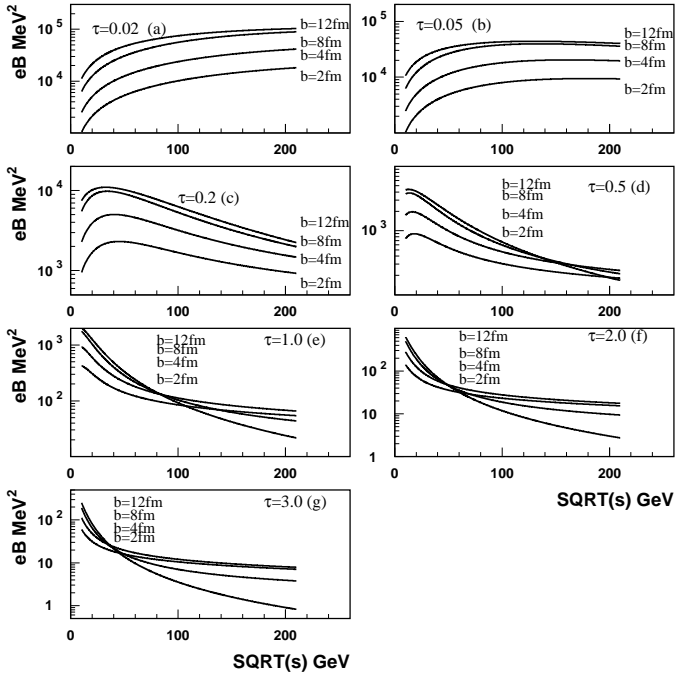}}
\caption{The dependencies of the magnetic field on collision energies of central of mass system less than 200 GeV at RHIC energy region at different impact parameters for
Au-Au collisions at different proper times for  $\tau$ =  0.02 fm (a), $\tau$ =  0.05 fm (b), $\tau$ =  0.2 fm (c),
$\tau$ =  0.5 fm (d), $\tau$ =  1.0 fm (e), $\tau$ =  2.0 fm (f) and $\tau$ =  3.0 fm (g), respectively.}
\label{fig4}
\end{figure}

The dependencies of the magnetic field on collision energies $\sqrt{s}$ at RHIC energy region at different impact parameters for Au-Au collisions
at different proper time $\tau$ are shown in Fig.4. It is found that the  dependencies of  magnetic field on collision energies
at different impact parameter are all on the rising trend at $\tau$ =  0.02 fm and $\tau$ =  0.05 fm as shown in Fig. 4(a, b).
Fig. 4(c)($\tau$ =  0.2 fm) shows that the magnetic
fields increase with the increasing of collision energies when $\sqrt{s}\leq30$ GeV, but then decrease with the increasing of collision energies when $\sqrt{s}>30$ GeV.
As the proper time $\tau$ increase to $\tau\geq0.5$ as shown in Figure 4(d, e, f, g),
the magnetic fields  decrease with the increasing of collision energies, and decrease more rapidly of more off-central collisions of  $b$ = 12  than the
more central collisions of  $b$ = 2 and 4 fm.

\begin{figure}[h!]
\centering \resizebox{0.5\textwidth}{!}{
\includegraphics{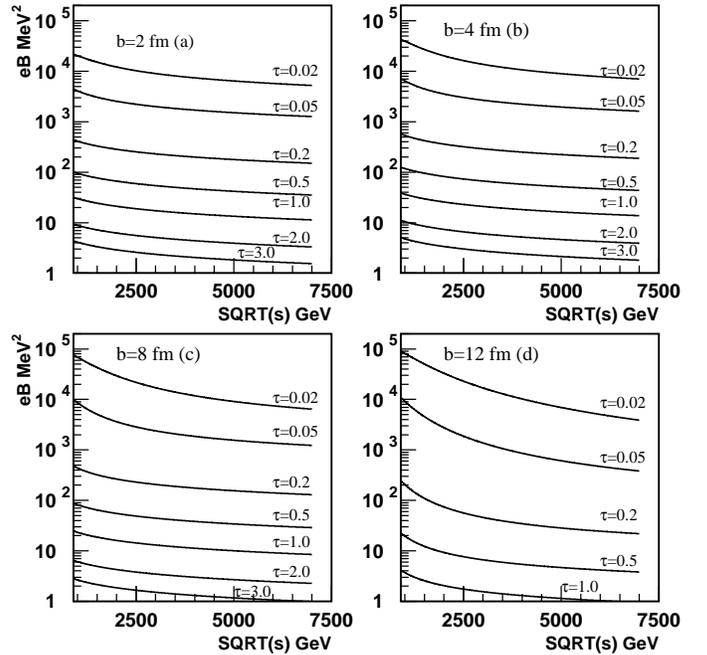}}
\caption{The dependencies of the magnetic field on collision energies of central of mass system of LHC energy region
of Pb-Pb collisions at different proper times at different
impact parameters for  $b$ =  2 fm (a),  $b$ =  4 fm  (b), $b$ =  8 fm (c) and
$b$ =  12 fm (d), respectively.}
\label{fig5}
\end{figure}

Figure 5 shows that the dependencies of the magnetic field on collision energies of central of mass system of LHC energy region
of Pb-Pb collisions at different proper times at different impact parameters for  $b$ =  2 fm,  $b$ =  4 fm, $b$ =  8 fm and
$b$ =  12 fm  with different proper time $\tau$. It is found that the magnitudes of magnetic fields decrease with increasing energy at LHC energy region,
and the speed of decreasing at large impact parameter such as $b$ = 12 fm (shown in Fig. (5d)) is more quickly than that of more central collision such as $b$ = 2 fm
(shown in Fig. 5(a)).  The magnetic field approach zero as $\tau \geq 1.0$ fm when $b$ = 12 fm.

\begin{figure}[h!]
\centering \resizebox{0.50\textwidth}{!}{
\includegraphics{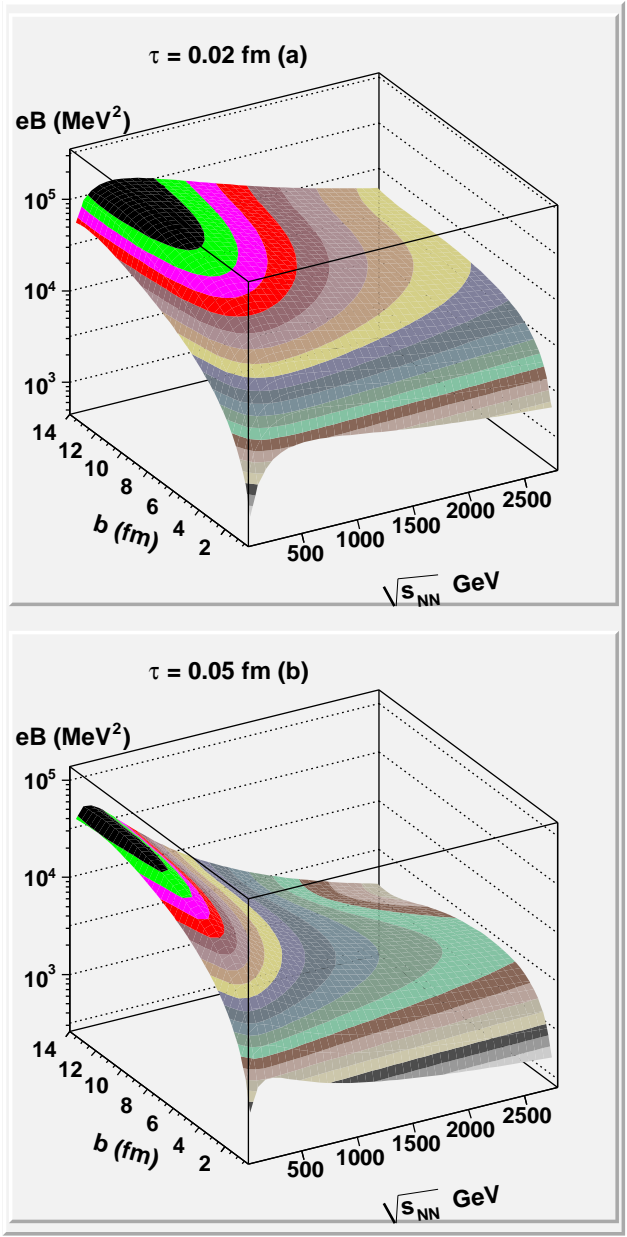}}
\caption{The dependencies of the magnetic field on collision energies of central of mass system and impact parameters
at different proper times for $\tau$ =  0.02 fm (a) and $\tau$ =  0.05 fm (b)}
\label{fig6}
\end{figure}

The dependencies of the magnetic field on collision energies of central of mass system and impact parameters
at different proper times for $\tau$ =  0.02 fm, 0.05 fm,  1.0 fm, 2.0 fm and 3.0 fm are shown in Fig. 6 and Fig. 7, respectively.
The three-dimensional schemes of different proper time show that the large magnetic fields are produced during small $\tau$, large
impact parameter and $\sqrt(s_{NN})\leq500$ GeV energy regions. The magnetic fields decrease with increasing collision energy during $\sqrt(s_{NN})\geq500$ GeV.
The maximum of magnetic field is $eB \simeq 50 MeV^{2}$ when $\tau = 2.0$ fm, but that of $\tau = 0.02$ fm is $eB \simeq 10^{5} MeV^{2}$.

\begin{figure}[h!]
\centering \resizebox{0.50\textwidth}{!}{
\includegraphics{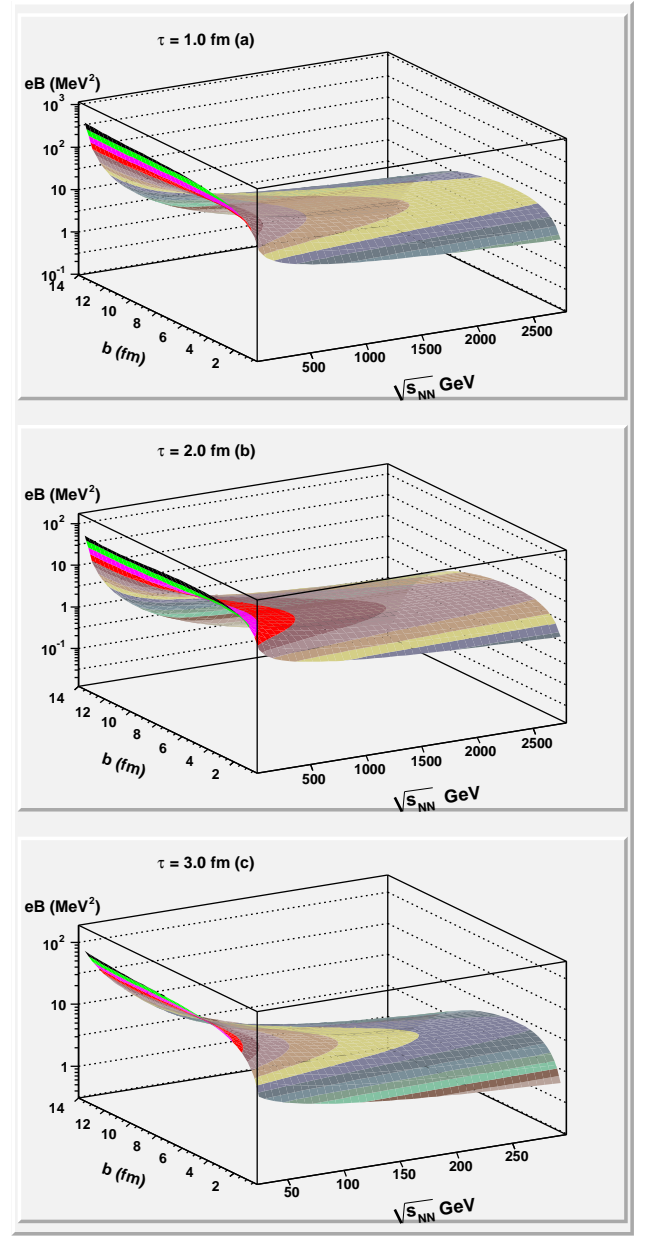}}
\caption{The dependencies of the magnetic field on collision energies of central of mass system and impact parameters
at different proper times for $\tau$ =  1.0 fm (a), $\tau$ =  2.0 fm (b) and $\tau$ =  3.0 fm (c), respectively}
\label{fig8}
\end{figure}

\section{Summary and Conclusion}
It is found that the relativistically moving heavy ions, typically with
large positive charges, carry strong magnetic (and electric)
fields with them. In the short moments before/during/after the impact of
two ions in non-central collisions, there is a very strong magnetic field in the
reaction zone. In fact, such a magnetic field is estimated to be of the
order of $m_{\pi}^{2}\approx 10^{18}$ Gauss, probably the strongest magnetic field in the present universe.
It appears at least during the very early stage of a heavy ion collision, there can be both
strong magnetic field and nonzero axial charge density in the created hot
matter.

It is suggested that off-central heavy-ion collisions can create strong transient
magnetic fields due to the fast, oppositely directed motion of two colliding
ions.  Thus,  a unique terrestrial
environment to study QCD in strong magnetic fields is provided in relativistic heavy-ion collisions. A strong magnetic
field can convert topological charge fluctuations in the QCD vacuum into global electric
charge separation with respect to the reaction plane. This so-called chiral magnetic effect may serve as a sign of the
local P and CP violation of QCD.

The feature of  magnetic fields at LHC $\sqrt{s_{NN}}$= 900, 2760 and 7000 GeV and at RHIC  $\sqrt{s_{NN}}$ = 62.4, 130 and 200 GeV
are systematically discussed. The dependencies of the features of magnetic fields on the collision energies, centralities and collision time are systematically investigated, respectively.

We show that an enormous magnetic field can indeed
be created in off-central heavy-ion collisions. The magnitude of the field is quite large, especially just after the collision, and decreases rapidly with time.
The drop velocity increases with the collision energy increase. It is shown that the magnitudes of magnetic fields decrease with increasing energy at LHC energy region,
and the speed of decreasing at large impact parameter such as $b$ = 12 fm  is more quickly than that of more central collision such as $b$ = 2 fm.

The  dependencies of the magnetic field on proper time and impact parameters for
at RHIC and LHC energy regions, respectively.  Comparing with that of RHIC energy region, we find that the magnitudes of the magnetic fields with
proper time fall more rapidly at LHC energy region. The variation characteristics of magnetic field with impact parameter at RHIC energy region are
different from that of LHC energy region.  The maximum position is located in the small proper time ($\tau\sim0.02 fm$), more off-central collisions of ($b\simeq 12 fm$) and
$\sqrt{s_{NN}}\sim200$ GeV. The maximum of magnetic field in our calculation is about $eB \simeq 10^{5} MeV^{2}$ when $\tau = 0.02$, $b\simeq 12 fm$ and $\sqrt{s_{NN}}\sim200$ GeV.

When surveying the dependencies of magnetic field on the proper time $\tau$, we finds that the magnitude of the magnetic field decrease with the increasing energy when $\sqrt{s_{NN}} \geq 200$ GeV.  The systematically research of magnetic field is consistent with the suggestion given by Ref.~\cite{lab24,lab25}
that the chiral magnetic effect  should either not change or should
decrease with increasing energy as long as a deconfined state of matter is formed in a heavy-ion collision.

\noindent \vskip1.0cm
{\bf Conflict of Interests}

\noindent The authors declare that there is no conflict of interests regarding the publication of this paper.

\section{Acknowledgments}
This work was supported in part by National Natural Science Foundation of
China (Grant No: 11375069, 11075061, 10975091£¬ 11075061 and 11221504),
 by the Ministry of Education of China under Grant No. 306022 ,
by the Programme of Introducing Talents of Discipline to Universities under Grant No. B08033 and
Key Laboratory foundation of Quark and Lepton Physics (Central China Normal University)(QLPL2014P01).

{}

\end{document}